\documentclass[twocolumn,notitlepage,prl,superscriptaddress,longtable, longbibliography]{revtex4-2}
\usepackage{mhchem}
\usepackage{amsthm}
\usepackage{amsmath}
\usepackage{amssymb}
\usepackage{mathdots}
\usepackage{graphicx}
\usepackage{mathrsfs}
\usepackage{longtable}
\usepackage{multirow}
\makeatletter

\@ifundefined{textcolor}{}
{%
 \definecolor{BLACK}{gray}{0}
 \definecolor{WHITE}{gray}{1}
 \definecolor{RED}{rgb}{1,0,0}
 \definecolor{GREEN}{rgb}{0,1,0}
 \definecolor{BLUE}{rgb}{0,0,1}
 \definecolor{CYAN}{cmyk}{1,0,0,0}
 \definecolor{MAGENTA}{cmyk}{0,1,0,0}
 \definecolor{YELLOW}{cmyk}{0,0,1,0}
}

\usepackage{braket}
\usepackage[backref=true,
bookmarksnumbered=true,
bookmarks=true,
bookmarksopen=true,
colorlinks=true,
citecolor=blue,
linkcolor=blue,
anchorcolor=green,
urlcolor=blue,unicode=false]{hyperref}

\renewcommand{\v}[1]{\ensuremath{\mathbf{#1}}} 

\newcommand{\abs}[1]{\left| #1 \right|} 

\let\baraccent=\= 
\renewcommand{\=}[1]{\stackrel{#1}{=}} 

\DeclareMathOperator{\diag}{diag}

\begin{document}
\title{Topological space-time crystal}
\author{Yang Peng}
\email{yang.peng@csun.edu}
\affiliation{Department of Physics and Astronomy, California State University, Northridge, Northridge, California 91330, USA}
\affiliation{Department of Physics, California Institute of Technology, Pasadena, California 91125, USA}

\begin{abstract}
We introduce a new class of out-of-equilibrium noninteracting topological phases, the topological space-time crystals. These are time-dependent quantum systems which do not have discrete spatial translation symmetries, but instead are invariant under discrete space-time translations.
Similar to the Floquet-Bloch systems, the space-time crystals can be described by a frequency-domain-enlarged
Hamiltonian, which is used to classify topologically distinct space-time crystals.  We show that these space-time
crystals can be engineered from conventional crystals with an additional time-dependent drive that behaves like a
traveling wave moving across the crystal. Interestingly, we are able to construct 1D and 2D examples of topological
space-time crystals based on tight-binding models which involve only one orbital, in contrast to the two-orbital minimal
models for any previously discovered static or Floquet topological phases with crystalline structures.
\end{abstract}

\maketitle
\emph{Introduction.}---
Symmetry has been shown to play an important role in topological classification of states of matter. For noninteracting fermionic systems in the presence of discrete spatial translation symmetry (i.e. crystals), the topological phase is characterized by the band structure topology, which is constrained by other co-existing symmetries, including the onsite ones \cite{Schnyder2008,Kitaev2009,Ryu2010,Teo2010,Chiu2016}, and possibly other crystalline symmetries from the crystal's space group \cite{Fu2011,Chiu2013,Slager2013,Shiozaki2014,Ando2015,Kruthoff2017,Bradlyn2017,Song2018,Song2019}.  

Such a symmetry-based topological classification scheme persists even when crystals are driven out of equilibrium, which opens the possibility of engineering desired topological properties with external knobs. As a paradigm, Floquet engineering, i.e. the control of quantum systems by time-periodic external fields turns out to be extremely powerful. For example, 
robust electron conducting channels can be brought at the boundary of an otherwise trivial two-dimensional insulator,
upon a circularly polarized irradiation or an alternating Zeeman field \cite{Oka2009,Inoue2010,Kitagawa2011,Lindner2011,Lindner2013}.
More generally, a complete classification of the Floquet topological insulators and superconductors when considering the 
onsite symmetries has been obtained in Ref.~\cite{Roy2017, Yao2017}, and the classification can be further enriched if more crystalline symmetries are taken into account \cite{Konstantinos2019,Yu2021}. 

When exploring the possible topological phases out of equilibrium, besides focusing on the same symmetries as in static situations,
space-time symmetries \cite{Morimoto2017,Xu2018}, which relate different parts of systems at different times, should also be considered, as they arise completely due to the new-added time dimension and have no analogue in an equilibrium setup. 
Indeed, certain space-time symmetries are shown to lead to a different topological classification from the one obtained using purely crystalline symmetries \cite{Peng2019,Peng2020,Chaudhary2020}. Yet, there are some limitations in these works because the space-time symmetries considered are constrained by requiring the underlying system being a periodically driven crystal, with separate discrete translation symmetries along spatial and temporal dimensions (which we refer to as the ``Floquet-Bloch'' scenario). 

\begin{figure}
    \centering
    \includegraphics[width=0.48\textwidth]{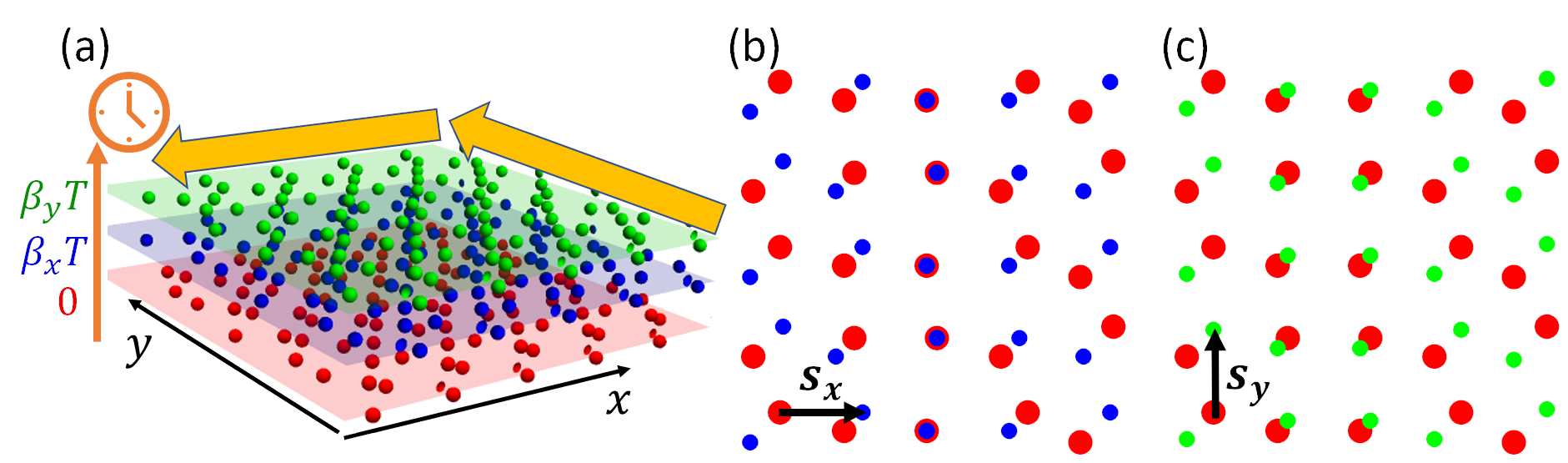}
    \caption{Illustration of a 2D topological space-time crystal in class A. (a) The vertical orange axis denotes the time coordinate. The 2D non-crystalline lattice system at three different times are shown in three colors. Besides being time-periodic with period $T$, the system is invariant under two additional discrete space-time translations: $(\boldsymbol{s_x},\beta_x T)$ and $(\boldsymbol{s_y},\beta_y T)$. The system hosts chiral boundary modes at quasienergy $\pi/T$, indicated in yellow arrows. (b),(c) Systems at different times are shown on top of each other (in different colors): $t=0$ with $t=\beta_x T$, and $t=0$ with $t=\beta_yT$, respectively. The two spatial translations are also shown. }
    \label{fig:illustration}
\end{figure}

It has been proposed in Ref.~\cite{Xu2018} that more space-time symmetries may exist if we remove the above mentioned constraint, and we can have the so-called "space-time crystals", which is characterized by the existence of multiple discrete translation symmetries in the combination of spatial and temporal dimensions. Then one natural question to ask is can we have topologically protected states in such systems? 
In this letter, we shall provide a definite answer to this question. An example of a 2D space-time crystal in class A supporting topological protected chiral edge modes is illustrated in Fig.~\ref{fig:illustration}.

\emph{Space-time crystals.}--- As first introduced in Ref.~\cite{Xu2018}, the Hamiltonian $H = \xi(\hat{\boldsymbol{p}}) + V(\boldsymbol{r},t)$ describes a particle moving in $(D+1)$-dimensional space-time, parametrized as $(\boldsymbol{r},t)$. Here, $\xi(\hat{\boldsymbol{p}})$ is the momentum ($\hat{\boldsymbol{p}}$) dependent kinetic energy, and the potential $V(\boldsymbol{r},t)$ is assumed to have $(D+1)$ linearly independent discrete space-time translation symmetries along $(\boldsymbol{s}_i,\tau_i)$ for $i=0,1,2,\dots,D$ as $V(\boldsymbol{r},t) = V(\boldsymbol{r}+\boldsymbol{s}_i,t+\tau_i)$, which generalizes the familiar Hamiltonian in the Floquet-Bloch case.

Because of the generalized Floquet-Bloch theorem \cite{Xu2018}, the time-dependent Schr\"odinger equation $(i\partial_t - H)\psi(\boldsymbol{r},t)=0$ admits solutions labeled by the quasimomentum $\boldsymbol{k}$ and quasienergy $\omega$, $\psi_{\boldsymbol{k},\omega}(\boldsymbol{r},t) = e^{i(\boldsymbol{k}\cdot\boldsymbol{r} -\omega t)}u_{\boldsymbol{k},\omega}(\boldsymbol{r},t)$, where $u_{\boldsymbol{k},\omega}(\boldsymbol{r},t)$ has the same space-time translation symmetries as the ones in $V(\boldsymbol{r},t)$.
Similar to the situation in a periodically driven crystal,  $\boldsymbol{k}$ and $\omega$ are defined up to adding reciprocal vectors of the form $(\boldsymbol{G}_j,\Omega_j)$ with $\boldsymbol{s}_i \cdot \boldsymbol{G}_j - \tau_i\Omega_j = 2\pi \delta_{ij}$, 
and thus should be restricted in the first (energy-momentum) Brillouin zone (BZ).  Moreover, the momentum $\boldsymbol{k}$ and energy $\omega$ are not independent and they must solve the following eigenvalue equation
\begin{equation}
    \left[\xi(\hat{\boldsymbol{p}}+\boldsymbol{k})+V(\boldsymbol{r},t) - i\partial_t \right]u_{\boldsymbol{k},\omega}(\boldsymbol{r},t) = \omega u_{\boldsymbol{k},\omega}(\boldsymbol{r},t). \label{eq:eigvalue_K}
\end{equation}

To find out the solution at $(\boldsymbol{k},\omega)$, one can solve the eigenvalues at each $\boldsymbol{k}$ in Eq.~(\ref{eq:eigvalue_K}), and identify solutions that are separated by multiple combination of reciprocal vectors $(\boldsymbol{G}_j,\Omega_j)$ in energy-momentum space. 
This can also be seen if we introduce $\tilde{u} = e^{i(\boldsymbol{G}_j\cdot \boldsymbol{r} - \Omega_j t)}u_{\boldsymbol{k},\omega}(\boldsymbol{r},t)$, 
then $\tilde{u}$ solves the same type of eigenvalue equation with substitution $\boldsymbol{k}\to \boldsymbol{k}-\boldsymbol{G}_j$, $\omega \to \omega-\Omega_j$,
and thus we can denote $\tilde{u} \equiv u_{\boldsymbol{k}-\boldsymbol{G}_j,\omega-\Omega_j}$.
Notice that equations similar to Eq.~(\ref{eq:eigvalue_K}) also exist when we simply have Floquet-Bloch systems, in which momentum and energy are separately identified upon addition of multiple reciprocal (to the spatial dimensions) vectors and the driving frequency, respectively. This similarity motivates us to think of band topology in a space-time crystal. 

\emph{Spectrum.}---To generalize the concepts of (Floquet) topological insulators and superconductors to the space-time crystals, we have to focus on systems with an energy gap in the spectrum. However, if there exist more than two nonzero $\Omega_j$s in the reciprocal vectors $(\boldsymbol{G}_j, \Omega_j)$, the spectrum will be dense and thus gapless. To see this, one can assume $(\boldsymbol{k},\omega)$ is a solution, and $\Omega_1$, $\Omega_2$ are two nonzero frequencies in the reciprocal vectors. It implies that all states labeled by $(\boldsymbol{k}+n\boldsymbol{G}_1 + m\boldsymbol{G}_2, \omega+n\Omega_1 + m\Omega_2)$ with $n,m\in\mathbb{Z}$ are also solutions. Since $\Omega_1$ and $\Omega_2$ are incommensurate (otherwise one can redefine reciprocal vectors to make one of them zero),  $n\Omega_1+m\Omega_2$ can be made arbitrarily close to zero with $\abs{n},\abs{m}\to \infty$, similar to the situation in the quasiperiodically driven case considered in Ref.~\cite{Peng2018}.  

Hence, in order to have a "gapped" space-time crystal, we will focus on the particular situations where there only exists one nonzero frequency.
In this case, the space-time discrete translation basis vectors can be parametrized as $(\boldsymbol{0},-T)$, and $(\boldsymbol{s}_j, \beta_j T)$ with $\beta_j \in (-\frac{1}{2},\frac{1}{2}]$, $j=1,2,\dots,D$. The corresponding reciprocal vectors in energy-momentum space are $(\sum_{j=1}^D \beta_j \boldsymbol{G}_j, \Omega)$ with $\Omega=2\pi/T$, and $(\boldsymbol{G}_j, 0)$ for $j=1,2,\dots, D$. This implies that the potential can be written as
\begin{equation}
    V(\boldsymbol{r},t) = \sum_{n \in \mathbb{Z}} e^{in(\sum_j \beta_j \boldsymbol{G}_j\cdot \boldsymbol{r} - \Omega t)} V_n (\boldsymbol{r}),
\end{equation}
where $V_n(\boldsymbol{r})=V_n(\boldsymbol{r}+\boldsymbol{s}_j)$ is spatially periodic in $D$ dimensions, for all $n\in \mathbb{Z}$ and $j=1,2,\dots,D$.
They satisfy $V_n(\boldsymbol{r}) = V_{-n}(\boldsymbol{r})$ because of hermiticity.

Similarly, we can write 
\begin{equation}
    u_{\boldsymbol{k},\omega}(\boldsymbol{r},t) = \sum_{n\in \mathbb{Z}} e^{in(\sum_j \beta_j\boldsymbol{G}_j \cdot \boldsymbol{r} - \Omega t)} u_{n,\boldsymbol{k},\omega}(r),
\end{equation}
and one can show that the function $u_{n,\boldsymbol{k},\omega}(\boldsymbol{r})$ satisfies
\begin{align}
        &\left[h_0\left(\boldsymbol{k}+m \delta \boldsymbol{k}\right)-m\Omega\right]u_{m,\boldsymbol{k},\omega}(\boldsymbol{r}) \nonumber\\
        + &\sum_{n\neq 0} V_n(\boldsymbol{r})u_{m-n,\boldsymbol{k},\omega}(\boldsymbol{r}) = \omega u_{m,\boldsymbol{k},\omega}(\boldsymbol{r}), \label{eq:extended_Floquet}
\end{align}
where $\delta \boldsymbol{k} = \sum_j \beta_j \boldsymbol{G}_j$, and $h_0(\boldsymbol{k}) = \xi(\hat{\boldsymbol{p}}+\boldsymbol{k}) + V_0(\boldsymbol{r})$ can be interpreted as a Bloch Hamiltonian describing a particle moving in the (spatially) periodic potential $V_0(\boldsymbol{r})$.

Let us denote the eigenstates of $h_0(\boldsymbol{k})$ as $u^{j}_{\boldsymbol{k}}$ at energy $\epsilon_{\boldsymbol{k}}^j$ (assuming $j=1,2,\dots,d$), and we can expand $u_{m,\boldsymbol{k},\omega} = \sum_j c_{m,\boldsymbol{k},\omega}^j u^j_{\boldsymbol{k}+m\delta \boldsymbol{k}}$. Thus, Eq.~(\ref{eq:extended_Floquet}) can be converted into an eigenvalue equation $\mathscr{H}(\boldsymbol{k})\boldsymbol{C}_{\boldsymbol{k},\omega} = \omega \boldsymbol{C}_{\boldsymbol{k},\omega}$, where we introduced the enlarged Hamiltonian in the frequency domain as
\begin{align}
&\mathscr{H}(\boldsymbol{k}) =\nonumber \\ 
&\left(\begin{array}{ccccc}
\ddots\\
 &h_0(\boldsymbol{k} - \delta\boldsymbol{k}) +\Omega & h_{1}^\dagger(\boldsymbol{k}- \delta\boldsymbol{k}) & h_{2}^\dagger(\boldsymbol{k}-\delta\boldsymbol{k})\\
 & h_{1}(\boldsymbol{k} - \delta\boldsymbol{k}) & h_0(\boldsymbol{k}) & h_{1}^\dagger(\boldsymbol{k}) \\
 & h_{2}(\boldsymbol{k} - \delta\boldsymbol{k}) & h_{1}(\boldsymbol{k}) & h_0(\boldsymbol{k}+\delta \boldsymbol{k})-\Omega\\
 &  &  &  & \ddots
\end{array}\right). \label{eq:enlarged_H}
\end{align}
Here, each diagonal block of $d$ dimensions corresponds to a particular frequency sector labeled by $m$. The column vector $\boldsymbol{C}_{\boldsymbol{k},\omega} = (\dots,\boldsymbol{c}_{-1,\boldsymbol{k},\omega}^T, \boldsymbol{c}_{0,\boldsymbol{k},\omega}^T,\boldsymbol{c}_{1,\boldsymbol{k},\omega}^T,\dots)^T$, with $\boldsymbol{c}_{m,\boldsymbol{k},\omega} = (c_{m,\boldsymbol{k},\omega}^1,\dots,c_{m,\boldsymbol{k},\omega}^d)^T$, consists all expansion coefficients accordingly. Here, the diagonal blocks are diagonalized with $h_0(\boldsymbol{k}) =  \diag{(\epsilon_{\boldsymbol{k}}^1,\dots,\epsilon_{\boldsymbol{k}}^d)}$ since eigenstates' basis is used. The off-diagonal blocks $\{h_n(\boldsymbol{k})\}$ are in general nondiagonal and their matrix elements are computed from $[h_n(\boldsymbol{k})]_{ij} = \bra{u^{i}_{\boldsymbol{k}+n\delta\boldsymbol{k}}}V_n\ket{u^{j}_{\boldsymbol{k}}}$. 

It is worth mentioning that $\mathscr{H}(\boldsymbol{k})$ looks very similar to the enlarged Hamiltonian for the
Floquet-Bloch problem in the frequency domain formulation \cite{Gomez2013}, except that the off-diagonal blocks no longer couple different Floquet sectors (indexed by $m$) at the same Bloch momentum $\boldsymbol{k}$, but rather allow mixing of states from different Floquet sectors with momentum differing in multiples of $\delta \boldsymbol{k}$. Thus, we expect these couplings can produce topologically protected energy gaps for space-time crystals. 

\emph{Space-time crystals from crystals.}--- The space-time crystals with gapped spectra can be realized from an ordinary crystal (including cold atoms in optical lattices) in equilibrium, described by the Hamiltonian $H_0 = \xi(\hat{\boldsymbol{p}}) + V_0(\boldsymbol{r})$, or simply the Bloch Hamiltonian $h_0(\boldsymbol{k})$. To make the system a space-time crystal, one further generates a traveling wave on the crystal lattices, by applying, for example,  an acoustic wave (or modulating the optical potential for cold atoms) at angular frequency $\Omega$ and wave vector $\delta \boldsymbol{k}$, which produces the term $\sum_{n\neq 0}e^{in(\delta \boldsymbol{k}\cdot \boldsymbol{r}- \Omega t)}V_n(\boldsymbol{r})$.  

When the static crystal with Bloch Hamiltonian $h(\boldsymbol{k})$ is described by a tight-binding model (namely using Wannier functions as a basis), the traveling wave potential generates an additional onsite and hopping terms which satisfy the space-time translation symmetries. This can be seen by introducing the Wannier function $\ket{w_{j\boldsymbol{R}}}$ at site $\boldsymbol{R}$ with orbital index $j$, for the crystalline Hamiltonian $H_0$. The hopping amplitude between neighboring sites (or onsite term for $\boldsymbol{R'}=\boldsymbol{R}$) is given by the matrix elements
\begin{equation}
    \bra{w_{i\boldsymbol{R}'}}H\ket{w_{j\boldsymbol{R}}}=[\tilde{h}^{0}(\boldsymbol{R}'- \boldsymbol{R})]_{ij}+\sum_{n\neq 0}[h^{n}(\boldsymbol{R}',\boldsymbol{R},t)]_{ij},
    \nonumber
\end{equation}
where $[\tilde{h}^{0}(\boldsymbol{R}'- \boldsymbol{R})]_{ij}$ are the static hopping/onsite terms which only depend on $\boldsymbol{R}'- \boldsymbol{R}$. The time-dependent hopping/onsite terms can be written in the following form \cite{suppl}
\begin{equation}
    [h^{n}(\boldsymbol{R}',\boldsymbol{R},t)]_{ij} = e^{in(\delta \boldsymbol{k}\cdot \boldsymbol{R}' - \Omega t)} [\tilde{h}^{n}(\boldsymbol{R}'- \boldsymbol{R})]_{ij},
    \label{eq:td_hopping}
\end{equation}
where $\tilde{h}^n$ is some function depending only on $\boldsymbol{R}'- \boldsymbol{R}$. Thus, we see the time-dependent hopping/onsite terms are invariant if we transform $\boldsymbol{R}'$, $\boldsymbol{R}$, and $t$ according to the space-time translations.

As shown in the supplemental material \cite{suppl}, the enlarged Hamiltonian introduced in Eq.~(\ref{eq:enlarged_H}) can also be derived using the Wannier functions. The tight-binding parameters $[\tilde{h}^{n}(\boldsymbol{R}'- \boldsymbol{R})]_{ij}$ are related to $\mathscr{H}(\boldsymbol{k})$ through
\begin{equation}
    h_n(\boldsymbol{k}) = \sum_{\boldsymbol{R}} e^{-i\boldsymbol{k}\cdot \boldsymbol{R}} \tilde{h}^n(\boldsymbol{R}), \quad (n\geq 0).
\end{equation}

\emph{Topological classification.}---The topological classification of the space-time crystals now boils down to the classification of the enlarged Hamiltonian $\mathscr{H}(\boldsymbol{k})$ at a particular energy gap, given a set of symmetries. 

Let $\mathscr{H}(\boldsymbol{k})$ be in one of the Altland-Zirnbauer (AZ) symmetry classes  \cite{Schnyder2008,Kitaev2009,Ryu2010,Teo2010,Chiu2016}, which are determined by the presence or absence of the time-reversal, particle-hole and chiral symmetries, defined according to
according to $\hat{\mathscr{T}}\mathscr{H}(\boldsymbol{k})\hat{\mathscr{T}}^{-1} = \mathscr{H}(\boldsymbol{k}_*-\boldsymbol{k})$, $\hat{\mathscr{C}}\mathscr{H}(\boldsymbol{k})\hat{\mathscr{C}}^{-1} = -\mathscr{H}(\boldsymbol{k}_*-\boldsymbol{k})$, and $\hat{\mathscr{S}}\mathscr{H}(\boldsymbol{k})\hat{\mathscr{S}}^{-1} = -\mathscr{H}(\boldsymbol{k})$, respectively.
Here, $\hat{\mathscr{T}}$ and $\hat{\mathscr{C}}$ are antiunitary operators, while $\hat{\mathscr{S}}$ is unitary. 
Note that we also generalized the definitions by allowing the time-reversal invariant momentum to be different from zero and located at $\boldsymbol{k}_*$. 
In other words, this means that under anti-linear operation, the momentum is reflected about $\boldsymbol{k}_*/2$ rather than zero. 
Thus, the classification of topological states at a given energy gap depends on the dimension of $\boldsymbol{k}$ and the AZ symmetry class of $\mathscr{H}(\boldsymbol{k})$, as in the case of topological insulators and superconductors. In principle, one can also consider other spatial symmetries instead of onsite symmetries considered above, which will enrich the topological classification. 

We are particularly interested in the topologically protected boundary modes at $\omega = \Omega/2 \mod \Omega$ with open boundary conditions, which indicates a nontrivial out-of-equilibrium topological state that has no analogue in the equilibrium scenario \cite{Rudner2013,Nathan2015}. In terms of the enlarged Hamiltonian, this often requires coupling between neighboring diagonal sectors in $\mathscr{H}(\boldsymbol{k})$. 

For simplicity, let us consider the harmonic driving protocols when $h_n(\boldsymbol{k})$ vanishes for $\abs{n}\geq 2$. Moreover, 
let us assume that the driving term $h_1(\boldsymbol{k})$ is small and there are energy overlap only between neighboring Floquet sectors along the diagonal in $\mathscr{H}(\boldsymbol{k})$, i.e $h_0(\boldsymbol{k})\simeq h_0(\boldsymbol{k}+\delta \boldsymbol{k})-\Omega \simeq -\Omega/2$ for some $\boldsymbol{k}$. With these assumptions, one can write $\mathscr{H}(\boldsymbol{k})\simeq \mathscr{H}_{\rm eff}(\boldsymbol{k}) - \rho_0\Omega/2$ for the low energy physics around $-\Omega/2$, where $\rho_0$ is the two-by-two identity matrix. The topological classification at the energy gap $-\Omega/2$ is then determined by the effective Hamiltonian
\begin{equation}
    \mathscr{H}_{\rm eff}(\boldsymbol{k}) = \left(\begin{array}{cc}
h_{0}(\boldsymbol{k}) & h_{1}^{\dagger}(\boldsymbol{k})\\
h_{1}(\boldsymbol{k}) & h_{0}(\boldsymbol{k}+\delta \boldsymbol{k})
\end{array}\right) + \frac{\Omega}{2}\rho_z,
\label{eq:effective_2by2}
\end{equation}
where we introduced the Pauli matrices $\rho_{x,y,z}$ corresponding to the Floquet sectors. 

\emph{1D model in class D.}---
It is known that 1D crystals in class D with a particle-hole symmetry $\hat{\mathscr{C}}^2=1$ have $\mathbb{Z}_2$ topological invariants and can support Majorana boundary modes. 
In fact, 1D space-time crystal in class D can also be topological at quasienergy $\Omega/2$, where nontrivial boundary modes exist.
Such a model can be constructed using the harmonic driving protocol described by the effective Hamiltonian $\mathscr{H}_{\rm eff}(\boldsymbol{k})$ in Eq.~(\ref{eq:effective_2by2}). Consider $h_0(k) = -2w\cos(ka)$ and $h_1(k) = \Delta \cos(ka+\delta k a/2)$, it is obvious that we have a particle-hole symmetry realized via $\rho_x \mathscr{H}_{\rm eff}(k)\rho_x = -\mathscr{H}_{\rm eff}(\pi-\delta k-k)^*$. 
As shown in the supplemental material~\cite{suppl}, this symmetry also exists for the full $\mathscr{H}(k)$. 

We first take $\delta k a = \pm \pi$, then $\mathscr{H}_{\rm eff}(k)$ takes the familiar form of the Hamiltonian for the Kitaev chain \cite{Kitaev2001}, a toy model for the topological superconductor which hosts boundary Majorana modes in an open chain for $4w\gtrsim \Omega$.  
On the other hand, even if $\delta k a$ deviates from $\pm \pi$, the particle-hole symmetry persists, and thus the boundary modes should not disappear as along as the bulk gap does not close.

\begin{figure}
    \centering
    \includegraphics[width=0.5\textwidth]{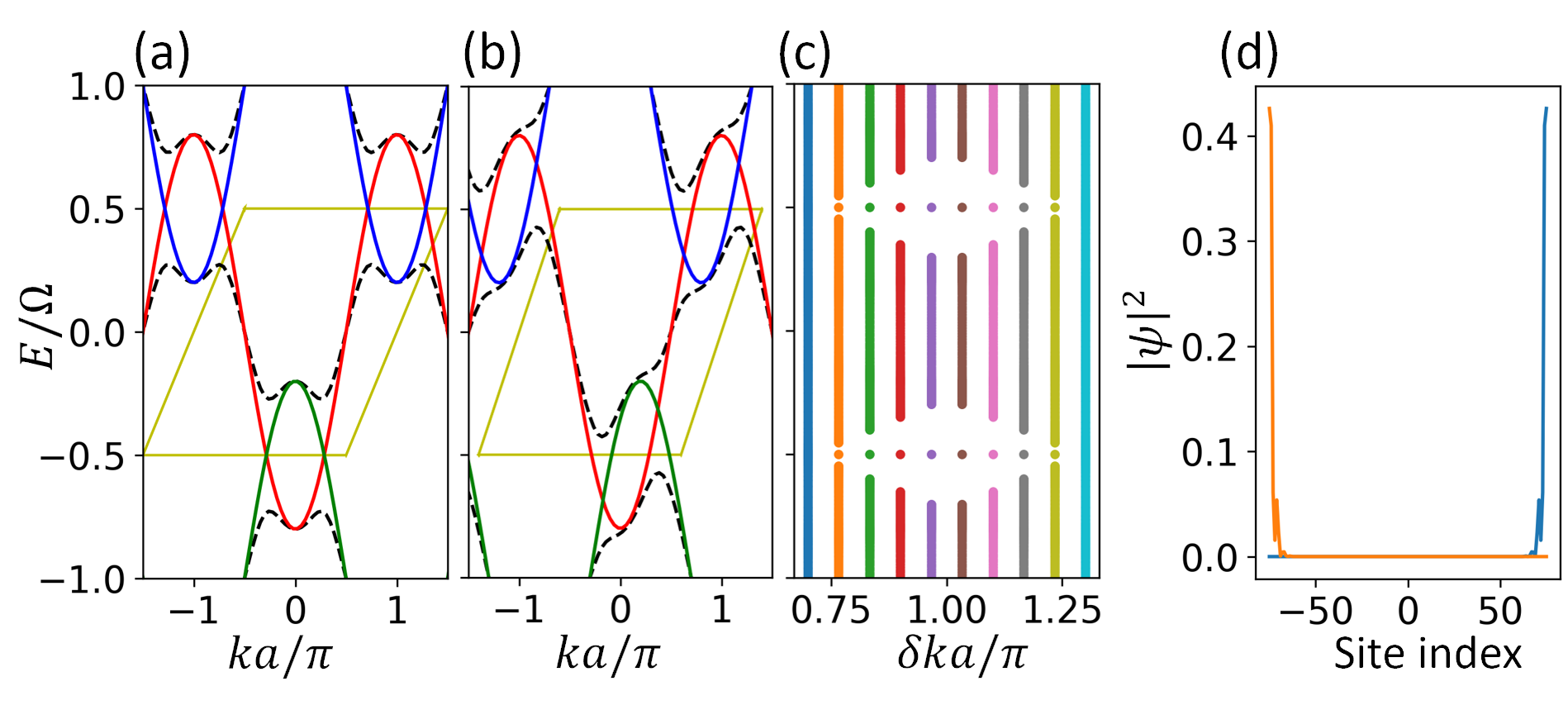}
    \caption{1D space-time crystal in class D. (a), (b) show the band structure of $\mathscr{H}(k)$ in black dashed lines, for $\delta k a/\pi = 1$ and $0.8$, respectively. The red, blue and green solid lines correspond to the diagonal $h_0$ in different Floquet sectors. The yellow parallelogram denotes the first energy-momentum BZ.  (c) Floquet spectra for an open chain at different $\delta k$. (d) Absolute value square of Majorana wave functions traced over Floquet sectors at $\delta k a = 0.8\pi$. The other parameters are $w/\Omega = 0.4$, $\Delta/\Omega = 0.3$, and $N_{\rm max}= 8$. The open chain contains 150 sites.}
    \label{fig:1Dmodel}
\end{figure}

Written in real space, the system is described by a one-orbital tight-binding model along $x$ direction with lattice spacing $a$
\begin{equation}
    H_{\rm 1D} = \sum_x\left[ (-w + f(x,t)) \psi_x^\dagger \psi_{x+a} + h.c.\right],
\end{equation}
where we have in the hopping parameter a space-time-translation invariant modulation $f(x,t) = \Delta \cos(\delta k
x-\Omega t+\frac{\delta ka}{2})$,  which satisfies $f(x,t)=f(x,t-2\pi/\Omega)=f(x+a,t+\delta k a/\Omega)$. 
The second-quantized operator $\psi_x^\dagger$ ($\psi_x$) creates (annihilates) an electron on site $x$.

The numerical results for the band structure of the full enlarged Hamiltonian $\mathscr{H}(k)$ (truncated up to $\pm
N_{\rm max}$ Floquet sector) are shown in Fig.~\ref{fig:1Dmodel}(a) and (b), for $\delta k a/\pi = 1$ and $0.8$,
respectively. With an open boundary condition, the the system loses its space-time translation symmetry and becomes a
Floquet non-crystalline chain. The spectra for such systems at different $\delta k$ are plotted in
Fig.~\ref{fig:1Dmodel}(c), which shows that the Majorana boundary modes at quasienergy $\Omega/2$ persist for a
considerable range of $\delta k a$ around $\pi$. In Fig.~\ref{fig:1Dmodel}(d), 
the absolute value square of the Majorana wave functions traced over all Floquet sectors are plotted for $\delta k a/\pi=0.8$.

\emph{2D model in class A.}---
It is known that 2D crystals in class A (no symmetries) have a $\mathbb{Z}$ topological invariant, the Chern number. 
Let us now construct a model for nontrivial 2D space-time crystals in class A, which has been illustrated in Fig.~\ref{fig:illustration}. 
For example, one can take $h_0(\boldsymbol{k}) = 2w(\cos(k_x a) + \cos(k_y a))$, and $h_1(\boldsymbol{k}) = \Delta[ \cos(k_x a+\delta k_x a/2) + i \cos(k_y a+\delta k_y a/2)]$. Choosing $\delta\boldsymbol{k}a = (\pi,\pi)$, the resulting $\mathscr{H}_{\rm eff}(k)$ becomes the half-BHZ model~\cite{Bernevig2006}, which can support chiral propagating edge modes at the boundary. 
Now if one considers $\delta \boldsymbol{k}$ away from $(\pi,\pi)$, as long as the bulk band gap does not close, the system should be in the same topological phase. 
\begin{figure}
    \centering
    \includegraphics[width=0.5\textwidth]{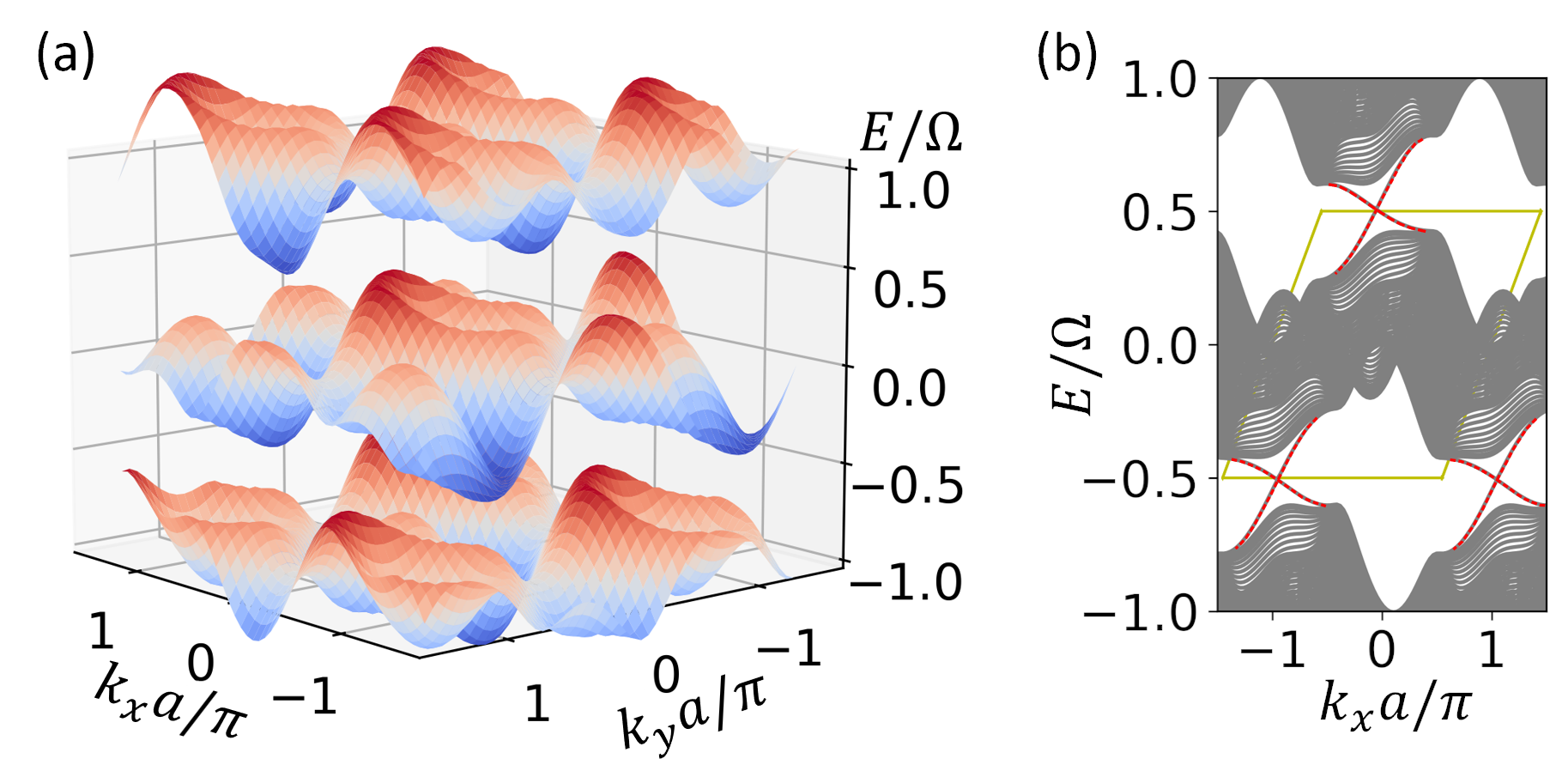}
    \caption{2D space-time crystal in class A. (a) Two dimensional band structure of $\mathscr{H}(\boldsymbol{k})$, for $\delta \boldsymbol{k} a = (0.9\pi,0.9\pi)$. (b) One dimensional band structure when periodic boundary condition is assumed only along $x$. The chiral edge modes inside the bulk gat at $\Omega/2$ are shown in red dotted lines. The energy-momentum BZ is denoted as the yellow parallelogram. The other parameters are $w/\Omega = 0.3$, $\Delta/\Omega = 0.2$, and $N_{\mathrm{max}} = 8$. The number of sites along $y$ is 50 in (b).}
    \label{fig:2Dmodel}
\end{figure}
\begin{figure}
  \centering
  \includegraphics[width=0.5\textwidth]{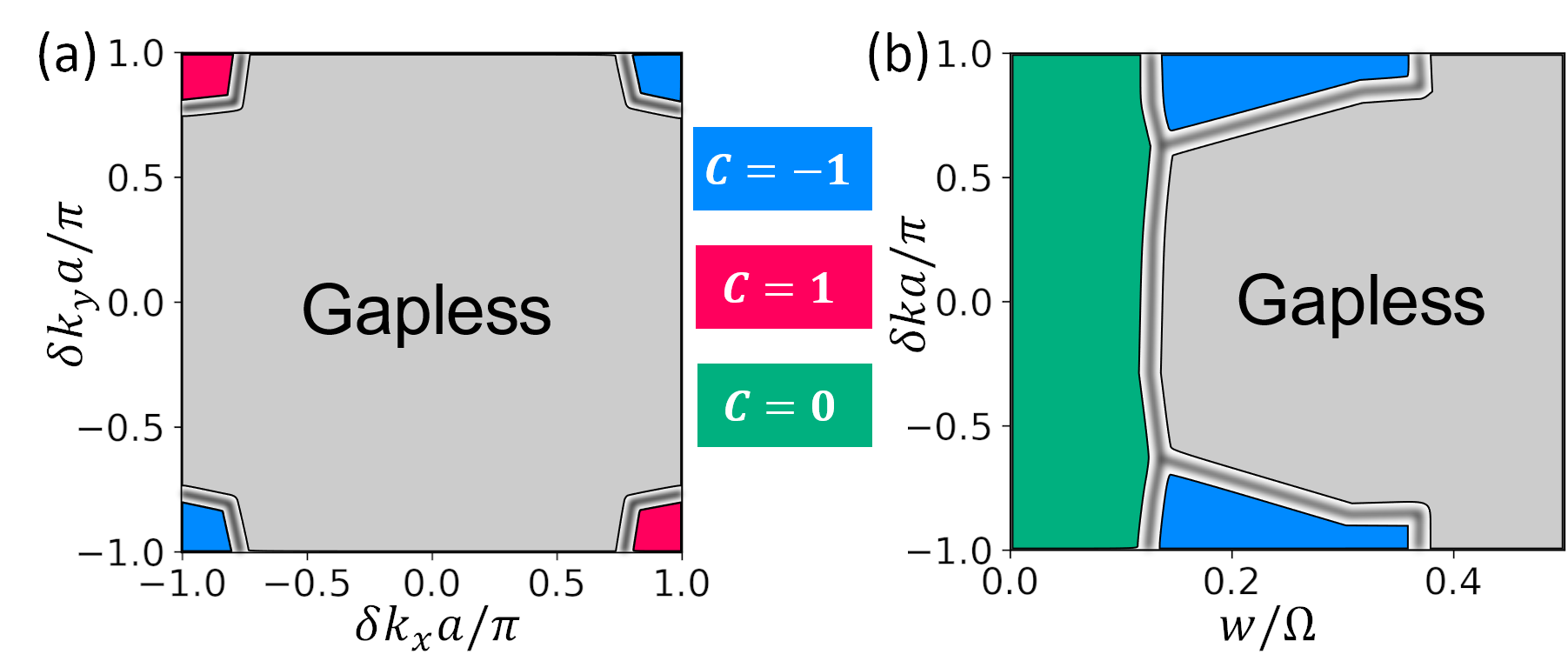}
  \caption{Phase diagram of 2D space-time crystal in class A. (a) Fixing $w/\Omega=0.3$, $\Delta/\Omega = 0.2$  while 
  varying $\delta k_x$ and $\delta k_y$, if the system is gapped at quasienergy $\Omega/2$, the topological invariant
  $C$ is computed.
  (b) Obtained by fixing $\Delta/\Omega=0.2$ while varying $w$ and $\delta k \equiv \delta k_x =\delta k_y$.
  }
  \label{fig:2Dphase}
\end{figure}

In real space, this system can be described by the following one-orbital tight-binding model on a square lattice with lattice constant $a$, whose Hamiltonian can be written as $H = H_0 + H_1$, where
\begin{equation}
    H_0 = w\sum_{\boldsymbol{R}}\left(\psi_{\boldsymbol{R}+a\hat{\boldsymbol{x}}}^\dagger \psi_{\boldsymbol{R}} + \psi_{\boldsymbol{R}+a\hat{\boldsymbol{y}}}^\dagger \psi_{\boldsymbol{R}} + h.c. \right),
\end{equation}
describes the static hopping between neighboring sites, while the second term 
\begin{align}
    H_1 &= \Delta\sum_{\boldsymbol{R}} \left[ \cos(\delta \boldsymbol{k}\cdot \boldsymbol{R} - \Omega t + \delta k_x /2)
    (\psi_{\boldsymbol{R}+a\hat{\boldsymbol{x}}}^\dagger \psi_{\boldsymbol{R}} + h.c.) \right. \nonumber  \\
  & \left. \sin(\delta \boldsymbol{k}\cdot \boldsymbol{R}  - \Omega t + \delta k_y /2)(\psi_{\boldsymbol{R}+a\hat{\boldsymbol{y}}}^\dagger \psi_{\boldsymbol{R}} + h.c.) \right]
\end{align}
corresponds to the time-dependent hopping with space-time translation symmetries. Here, $\psi^\dagger$ and $\psi$ denote the creation and annihilation operators. 

In Fig.~\ref{fig:2Dmodel} (a), we show a two-dimensional band structure of $\mathscr{H}(\boldsymbol{k})$ at $w/\Omega = 0.3$ and $\Delta/\Omega = 0.2$ at $\delta \boldsymbol{k} a/\pi = (0.9,0.9)$, which is expected to be gapped at energy $\Omega/2\mod \Omega$. Assuming open boundary condition along $y$ and periodic boundary condition along $x$, we lost one of the space-time translation symmetries and the enlarged Hamiltonian will only depend on one of the Bloch momentum, $k_x$. 
Diagonalization of it at each $k_x$, we see the subgap chiral modes around $\Omega/2$ in Fig.~\ref{fig:2Dmodel} (b), as expected.
In the supplemental material~\cite{suppl}, we show that these chiral modes are stable against weak disorder, similar to
the situation in strong topological insulators. 

In Fig.~\ref{fig:2Dphase} (a), the phase diagram as a function of $\delta k_x$ and $\delta k_y$ at fixed $w/\Omega = 0.3$
and $\Delta/\Omega = 0.2$ is shown. Apart from a large gapless region, the quasienergy spectrum is found to be gapped at
$\Omega/2$ (mod $\Omega$), at which the topological invariant $C$ can be computed by summing over the Chern numbers of
all bands below the gap. Note that $C$s computed from the enlarged Hamiltonian $\mathscr{H}(\boldsymbol{k})$
at every gapped energy $\Omega/2 + n\Omega$, with
$n\in\mathbb{Z}$, are all identical, due to the fact that the spectrum is shifted downward by $\Omega$ if one performs 
$\boldsymbol{k}\to\boldsymbol{k}+\delta\boldsymbol{k}$. Numerically, with a finite truncation parameter
$N_{\mathrm{max}}$, this means a nonzero Chern number can only be carried by the top/bottom-most bands. 
In Fig.~\ref{fig:2Dphase} (b), the phase diagram as a function of $w$ and $\delta k\equiv \delta k_x = \delta k_y$, at
fixed $\Delta/\Omega = 0.2$ is shown. 

\emph{Conclusion.}---
In this work, we extend the non-interacting topological phases to the scenario of space-time crystals, which are non-equilibrium quantum systems beyond the Floquet paradigm. Particularly, we focused on only a subset of space-time translation symmetries such that the bulk quasienergy spectrum can be gapped. We introduced the enlarged Hamiltonian, similar to the one used in the Floquet case, for the space-time crystal. Thus, the topological classification of the space-time crystal becomes the classification of the enlarged Hamiltonian, a static Bloch-like Hamiltonian. 
We showed that such space-time crystals can be engineered from conventional crystals with an additional time-dependent drive that behaves like a traveling wave moving across the crystal.  
 We further construct tight-binding models of topological space-time crystals in 1D and 2D to illustrate the general principle. 

The topological space-time crystals that support boundary modes at quasienergy $\Omega/2\mod\Omega$ are intrinsic non-equilibrium topological phases as they have no static analogue. More interestingly, the topological nontrivial systems can even be realized in one-orbital tight-binding models, whereas this is impossible for both static and Floquet noninteracting topological phases in which at least two orbitals are required. This would in some sense make topological space-time crystals experimentally even more accessible with a reduced degrees of freedom. 

Finally, it is worth mentioning that there are a few directions to generalize the results of this work. 
Although the construction of models in other AZ symmetry classes is straightforward, at least in the case of harmonic driving protocols as we can use the simple effective 2-by-2 Hamiltonian in Eq.~(\ref{eq:effective_2by2}), 
the classification with additional non-onsite space-time symmetries from a space-time group \cite{Xu2018} will require additional work. 
Another direction is to remove the restriction to the space-time translation symmetries considered in this work, and this will allow for dense quasienergy spectrum, as in a time-quasiperiodic system \cite{Peng2018}. 
Further more, here we only considered noninteracting space-time crystals. The exploration of topological phases that are stable in the presense of electron interactions will be left for future work. 
On the practical side, it is important to have experimentally feasible ways to detect the boundary modes in topological
space-time crystals. One possibility would rely on tunneling into the sample boundary at bias voltage at $\Omega/2$,
as proposed for the Floquet-Bloch systems \cite{Chaudhary2020,PengC2021}. 
Particularly, the tunneling conductance at the edge will be nonzero due to the spectral weight of the topological modes
\cite{Fregoso2014},  whereas it is vanishing when the tunneling happens inside the bulk. We numerically calculate
this spectral weight for the 1D and 2D models in the supplemental material \cite{suppl}.  

\emph{Acknowledgement.}--- 
Y.P. acknowledges support from the startup fund from California State University, Northridge. 
Y.P. is grateful for the helpful discussions with Gil Refael and Frederik Nathan.

%

\newpage
\begin{widetext}
\section*{Supplemental material}

\section*{Space-time crystal in tight-binding models}
In this section, let us derive the tight-binding representation of space-time crystals, based on Wannier functions in the crystal for a set of relevant energy bands, before the space-time translation invariant traveling wave is applied. We consider an isolated group of $d$ consecutive
Bloch bands $\{\phi^j_{\boldsymbol{k}}(\boldsymbol{r}) =
e^{i\boldsymbol{k}\cdot\boldsymbol{r}}u^j_{\boldsymbol{k}}(\boldsymbol{r}); j=1,2,\dots,d\})$ that do not become
degenerate with any lower or higher band anywhere in the Brillouin zone, such as the set of occupied valence bands in
insulators. Here $u^j_{\boldsymbol{k}}(\boldsymbol{r})$ is the eigenstates of the Bloch Hamiltonian $h_0(\boldsymbol{k})$ defined in the main text, and it has the translation symmetries of the crystal. 

The Wannier functions are constructed such that they span the same Hilbert space as the Bloch bands $\phi^j_{\boldsymbol{k}}(\boldsymbol{r})$, with $j=1,2,\dots,d$. There are $d$ Wannier functions centered at each lattice site $\boldsymbol{R}$, defined as
\begin{equation}
    w_{j\boldsymbol{R}}(\boldsymbol{r}) = \frac{1}{\sqrt{N}}\sum_{\boldsymbol{k}}e^{-i\boldsymbol{k}\cdot\boldsymbol{R}} \tilde{\phi}_{\boldsymbol{k}}^j(\boldsymbol{r}),
\end{equation}
where $N$ is the number of unit cells, and $\{\tilde{\phi}^j_{\boldsymbol{k}}(\boldsymbol{r})\}$ are a set of Bloch-like functions that are smooth in $\boldsymbol{k}$ everywhere in the Brillouin zone, and they are related to the energy eigenstates via a unitary transformation as
\begin{equation}
    \tilde{\phi}^j_{\boldsymbol{k}}(\boldsymbol{r}) = \sum_{i}U_{ij}(\boldsymbol{k})\phi^i_{\boldsymbol{k}}(\boldsymbol{r}). 
\end{equation}

Consider the Hamiltonian 
\begin{equation}
    H = \xi(\hat{\boldsymbol{p}}) + V_0(\boldsymbol{r}) + \sum_{n\neq 0 }e^{in(\delta\boldsymbol{k}\cdot \boldsymbol{r} - \Omega t)}V_n(\boldsymbol{r}),
\end{equation}
where the time-independent Hamiltonian $H_0 = \xi(\hat{\boldsymbol{p}}) + V_0(\boldsymbol{r})$ has eigenstates $\phi^j_{\boldsymbol{k}}(\boldsymbol{r})$, which are used to construct the Wannier functions.  
The Hamiltonian matrix elements in terms of Wannier functions can be written as
\begin{equation}
    \bra{w_{i\boldsymbol{R}'}}H\ket{w_{j\boldsymbol{R}}} = \bra{w_{i\boldsymbol{R}'}}H_0\ket{w_{j\boldsymbol{R}}} +  
    \sum_{n\neq 0}\bra{w_{i\boldsymbol{R}'}}e^{in(\delta\boldsymbol{k}\cdot \boldsymbol{r} - \Omega t)}V_n(\boldsymbol{r})\ket{w_{j\boldsymbol{R}}},
\end{equation}
where the first term $[\tilde{h}^{0}(\boldsymbol{R}'- \boldsymbol{R})]_{ij}\equiv\bra{w_{i\boldsymbol{R}'}}H_0\ket{w_{j\boldsymbol{R}}}$ corresponds to the static hopping/onsite terms in the tight-binding description of the crystal. 
Inside the summation of the second term, we have
\begin{align}
    \bra{w_{i\boldsymbol{R}'}}e^{in(\delta\boldsymbol{k}\cdot \boldsymbol{r} - \Omega t)}V_n(\boldsymbol{r})\ket{w_{j\boldsymbol{R}}}
    &= \int d\boldsymbol{r} w_{i\boldsymbol{R}'}^*(\boldsymbol{r}) e^{in(\delta\boldsymbol{k}\cdot \boldsymbol{r} - \Omega t)}V_n(\boldsymbol{r}) w_{j\boldsymbol{R}}(\boldsymbol{r})\nonumber \\
    & = \int d\boldsymbol{r} e^{in(\delta\boldsymbol{k}\cdot \boldsymbol{r}-\Omega t)} w_{i,\boldsymbol{0}}^*(\boldsymbol{r}-\boldsymbol{R}')V_n(\boldsymbol{r})w_{j,\boldsymbol{0}}(\boldsymbol{r}-\boldsymbol{R}) \nonumber \\
    & = e^{in (\delta \boldsymbol{k} \cdot \boldsymbol{R}' - \Omega t)} \int d\boldsymbol{r} w_{i,\boldsymbol{0}}^*(\boldsymbol{r})V_n(\boldsymbol{r})w_{j,\boldsymbol{0}}(\boldsymbol{r}+\boldsymbol{R}'-\boldsymbol{R})e^{in\delta\boldsymbol{k}\cdot\boldsymbol{r}},
\end{align}
where we have used the fact that the Wannier functions $w_{i,\boldsymbol{R}}(\boldsymbol{r}) = w_{i,\boldsymbol{0}}(\boldsymbol{r-\boldsymbol{R}})$, and $V_n(\boldsymbol{r}) = V_n(\boldsymbol{r}+\boldsymbol{R})$. If we define
\begin{equation}
    [\tilde{h}^{n}(\boldsymbol{R}'- \boldsymbol{R})]_{ij} = \int d\boldsymbol{r} w_{i,\boldsymbol{0}}^*(\boldsymbol{r})V_n(\boldsymbol{r})w_{j,\boldsymbol{0}}(\boldsymbol{r}+\boldsymbol{R}'-\boldsymbol{R})e^{in\delta\boldsymbol{k}\cdot\boldsymbol{r}},
\end{equation}
then we obtain Eq.~(6) in the main text. 

We can introduce the annhilation operator $\psi_{j,\boldsymbol{R}}$ for the $j$th Wannier function at site $\boldsymbol{R}$, and group $\{\psi_{j,\boldsymbol{R}}\}$ for $j=1,2,\dots,d$ into a column vector as $\boldsymbol{\psi}_{\boldsymbol{R}}$, and introduce $\boldsymbol{\psi}^\dagger_{\boldsymbol{R}}$ as the corresponding creation operators (in row vectors).  Thus, we obtained a space-time crystal described by the following tight-binding Hamiltonian
\begin{equation}
    H = \sum_{\boldsymbol{R}',\boldsymbol{R}}\boldsymbol{\psi}_{\boldsymbol{R}'}^\dagger\left[\tilde{h}^{0}(\boldsymbol{R}'- \boldsymbol{R}) + \sum_{n\neq0} e^{in (\delta \boldsymbol{k} \cdot \boldsymbol{R}' - \Omega t)} \tilde{h}^{n}(\boldsymbol{R}'- \boldsymbol{R})\right]\boldsymbol{\psi}_{\boldsymbol{R}}.
    \label{eq:H_tight_binding}
\end{equation}

For periodic boundary condition, we can write $\boldsymbol{\psi}_{\boldsymbol{R}} = 1/\sqrt{N}\sum_{\boldsymbol{k}} e^{i\boldsymbol{k}\cdot\boldsymbol{R}} \boldsymbol{a}_{\boldsymbol{k}}$ and $\tilde{h}^{n}(\boldsymbol{R}) = 1/N \sum_{\boldsymbol{q}}e^{i\boldsymbol{q}\cdot\boldsymbol{R}} h_n(\boldsymbol{q})$, then we have 
\begin{equation}
    H  = \sum_{\boldsymbol{k}}\left[ \boldsymbol{a}^\dagger_{\boldsymbol{k}} h_0(\boldsymbol{k}) \boldsymbol{a}_{\boldsymbol{k}} 
    + \sum_{n\neq 0} \boldsymbol{a}^\dagger_{\boldsymbol{k}+n\delta\boldsymbol{k}} h_n(\boldsymbol{k}) \boldsymbol{a}_{\boldsymbol{k}} e^{-in\Omega t} \right].
\end{equation}

The operator $K(t) = H(t) - i\partial_t$ can be Block diagonalized if we introduce $\boldsymbol{A}_{\boldsymbol{k}}(t) = (\dots, e^{i\Omega t} \boldsymbol{a}_{\boldsymbol{k}-\delta\boldsymbol{k}}^T, \boldsymbol{a}_{\boldsymbol{k}}^T,e^{-i\Omega t}\boldsymbol{a}_{\boldsymbol{k}+\delta\boldsymbol{k}}^T,\dots)^T$. This leads to 
\begin{equation}
    K(t) =  \sum_{\boldsymbol{k}} \boldsymbol{A}_{\boldsymbol{k}}^\dagger(t) \left[\mathscr{H}(\delta\boldsymbol{k},\boldsymbol{k}) - i\partial_t\right] \boldsymbol{A}_{\boldsymbol{k}}(t),
\end{equation}
where we obtain the effective enlarged \emph{time-independent} Hamiltonian $\mathscr{H}(\boldsymbol{k})$, which we have derived using the Bloch functions in the main text. 

\subsection{Particle-hole symmetries in $\mathscr{H}(k)$}
In the main text, we have shown that the effective Hamiltonian $\mathscr{H}_{\rm eff}(k)$ for the 1D model proposed in class D has a particle-hole symmetry realized via $\rho_x \mathscr{H}_{\rm eff}(k) \rho_x = -\mathscr{H}_{\rm eff}(\pi - \delta k -k)^*$. 
Here, we generalize this result by showing that any truncation of the frequency-domain-enlarged Hamiltonian $\mathscr{H}(k)$ has a particle-hole symmetry with respect to the energy, say $-\Omega/2$, as long as the truncation is symmetric about this energy. 

To make a truncation symmetric around $-\Omega/2$, we write
\begin{equation}
    \mathscr{H}(k)\simeq\mathscr{H}_{N}(k)-\mathbb{I}\Omega/2
\end{equation}
where the $2N$-by-$2N$ matrix
\begin{equation}
\mathscr{H}_{N}(k)=\left(\begin{array}{cccc}
h_{0}(k-(N-1)\delta k)+(N-\frac{1}{2})\Omega & h_{1}^{\dagger}(k-(N-1)\delta k)\\
h_{1}(k-(N-1)\delta k) & \ddots & \ddots\\
 & \ddots & \ddots & h_{1}^{\dagger}(k+(N-1)\delta k)\\
 &  & h_{1}(k+(N-1)\delta k) & h_{0}(k+N\delta k)-(N-\frac{1}{2})\Omega
\end{array}\right)
\end{equation}
and $\mathbb{I}$ is the identity matrix of the same size of $\mathscr{H}_{N}(k)$.
For $h_0(k) = -2w\cos(ka)$ and $h_1(k) = \Delta \cos(ka+\delta k a/2)$, one can easily verify that $\mathscr{H}_N(k)$ has a particle-hole symmetry realized via $X_N \mathscr{H}_N(k) X_N = -\mathscr{H}_N(\pi - \delta k - k)^*$, with the $2N$ by $2N$ matrix
\begin{equation}
    X_{N}=\left(\begin{array}{ccc}
 &  & 1\\
 & \iddots &\\
1 & &
\end{array}\right).
\end{equation}

\section*{Signature of topological edge states in tunneling conductance}
In this section, we will show that the topologically protected edge modes in space-time crystals can be probed by
measuring tunneling conductance, via either a scan tunneling microscope (STM) tip or a lead in a quantum transport
setup. The existence of edge quasienergy eigenstates will produce a nonzero conductance~\cite{Fregoso2014} between the edge of the
space-time crystal, and the STM tip (or lead).

In the following, for concreteness, let us focus on the STM-type setup.
Let us consider a (movable) tip is tunnel-coupled to the sample (the space-time crystal) at a location $\boldsymbol{r}$. The sample is
further weakly connected to a bath, which may locate beneath a layer of insulating substrate. 
Let us denote the tunneling rate between the sample and the tip (bath) as $\gamma_T$ ($\Gamma_B$).
It has been shown in Ref.~\cite{Fregoso2014} that in the tunneling regime when $\Gamma_T \ll \Gamma_B \ll$ the band width of the space-time crystal, 
the DC conductance at bias voltage $V$ can be written as
\begin{equation}
  G(V) \propto \frac{e^2\Gamma_T}{h}\mathrm{Im}[G_0^r(eV)]_{\boldsymbol{r},\boldsymbol{r}},
\end{equation}
where $G_n^r(\omega)$ is the sample's Floquet retarded Green function evaluated at spatial point $\boldsymbol{r}$, defined as
\begin{equation}
  G^r_n(\omega) = \frac{1}{T}\int_0^T dt\, \int_{-\infty}^{\infty} d\tau e^{in\Omega t}e^{i\omega \tau} G^r(t,t-\tau),
\end{equation}
with the retarded Green function written in terms of fermion creation and annihilation operators
\begin{equation}
  [G^r(t,t')]_{\boldsymbol{r},\boldsymbol{r}'} = -i\theta(t-t')\braket{\{\psi_{\boldsymbol{r}}(t),
  \psi_{\boldsymbol{r}'}^\dagger(t')\}}.
\end{equation}

The $G_0^r(\omega)$ can be computed from,
\begin{equation}
  G_0^r(\omega) = \sum_n \frac{\abs{\phi_n^{(0)}(\boldsymbol{r})}^2}{\omega - \epsilon_n + i\gamma_B},
\end{equation}
where $\varepsilon_n$ is quasienergy obtained by solving the eigenvalue problem for the enlarged Hamiltonian (written in coordinate space),
\begin{equation}
  \left(\begin{array}{ccccc}
    \ddots & \ddots\\
    \ddots & h_{0}+\Omega & h_{1}^{\dagger} & \ddots\\
    & h_{1} & h_{0} & h_{1}^{\dagger} & \ddots\\
    & \ddots & h_{1} & h_{0}-\Omega\\
    &  &  & \ddots & \ddots
  \end{array}\right)\left(\begin{array}{c}
    \vdots\\
    \phi_{n}^{(-1)}\\
    \phi_{n}^{(0)}\\
    \phi_{n}^{(1)}\\
    \vdots
  \end{array}\right)=\epsilon_{n}\left(\begin{array}{c}
    \vdots\\
    \phi_{n}^{(-1)}\\
    \phi_{n}^{(0)}\\
    \phi_{n}^{(1)}\\
    \vdots
  \end{array}\right).
\end{equation}

\begin{figure}[h]
  \centering
  \includegraphics[width=0.8\textwidth]{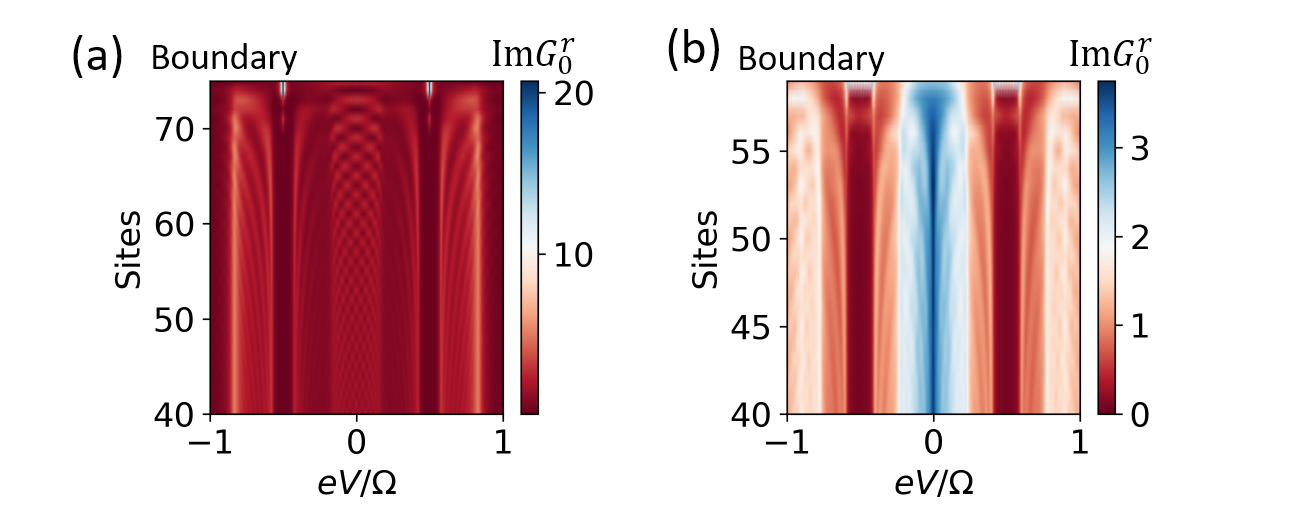}
  \caption{(a). Imaginary part of $G_0^r$ (in arbitrary unit) as a function of energy $eV$ at different spatial
  locations, in the 1D model of class D.
  We took the same parameters that generate Fig. 2(b) of the main text. (b) Same for the 2D model of class A, where we
  assume periodic boundary condition along $x$ and the vertical axis denotes the position along $y$ direction, at a
  fixed $x$. We took the same parameters that generate Fig. 3 of the main text. Additionally, we took $\gamma_B/\Omega =
  10^{-2}$. }
  \label{fig:spectral_function}
\end{figure}

In Fig.~\ref{fig:spectral_function}, we show the imaginary part of $G_0^r$ as a function of energy, or bias voltage
$eV$, at different spatial
locations of the sample, for both the 1D model of class D and the 2D model of class A introduced in the main text. 
It can be seen that the appearance of the quasienergy edge modes can be seen as nonzero conductance measured via
tunneling into the boundary of the sample (top of the figures), whereas the conductance drops to zero if the tunneling
is inside the bulk (bottom of the figures). 

\section*{Stability of 2D class A space-time crystal}
Like the topological edge modes in a strong topological insulator, spatial disorder that weakly breaks the space-time
translation symmetry cannot distroy the edge modes in space-time crystals, as long as the disorder does not break the
onsite AZ symmetries and the bulk gap does not close. For class A models in particular, as there are no additional AZ
symmetries, the topologically protected edge modes will be very stable against an arbitrary weak perturbation. 

\begin{figure}[h]
  \includegraphics[width=0.8\textwidth]{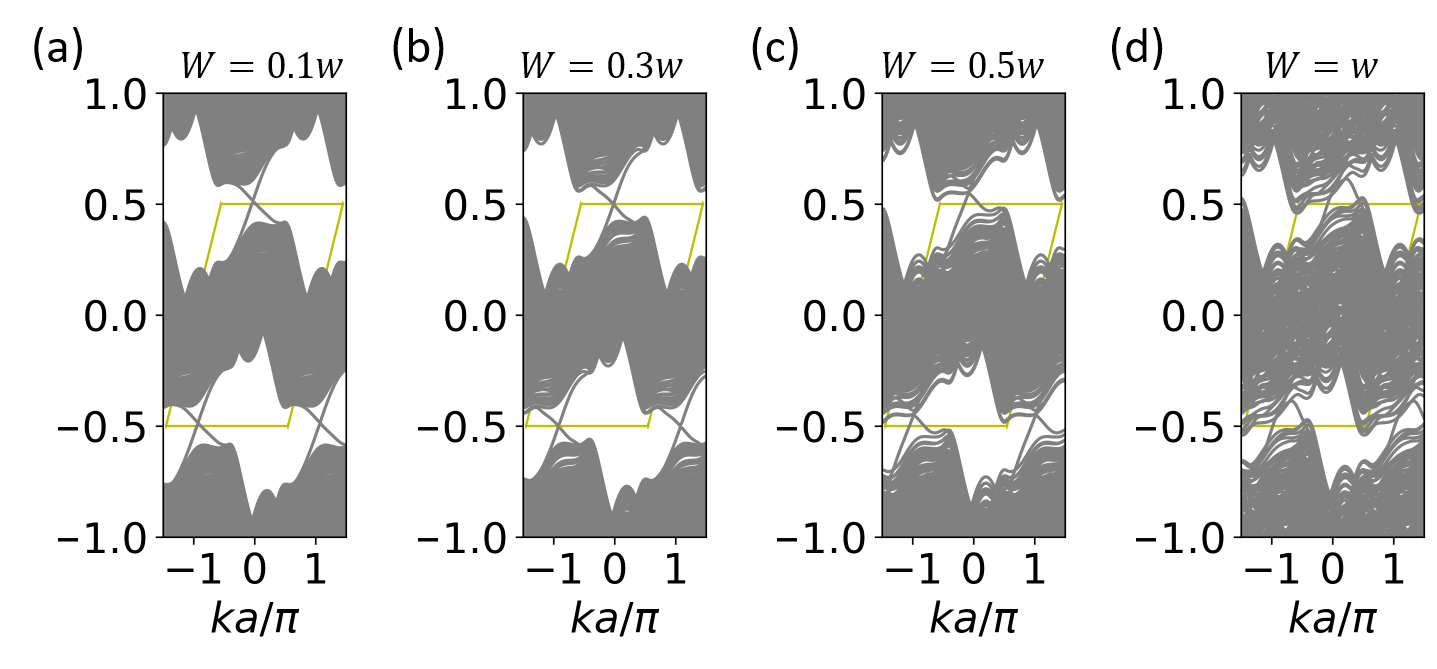}
  \caption{One dimensional band structure for the 2D model when periodic boundary condition
is assumed only along $x$, with the same parameters that generate the Fig.~3 (b) of the main text. In addition, we
add onsite random potentials of strength (a) $W=0.1w$, (b) $W=0.3w$, (c) $W=0.5w$, (d) $W=w$. }
\label{fig:disorder}
\end{figure}

To illustrate this, let us consider the 2D model introduced in the main text. For simplicity, let us take the periodic
boundary condition along $x$ and add to Eq.~(10) of the main text an independently identically distributed (i.i.d.) onsite random potential along the
$y$ direction, drawn from uniformly from the interval $[-W,W]$. In Fig.~\ref{fig:disorder}, we plot the 1D band
structure along $k_x$ with different disorder strength $W$, with other parameters same as the ones for Fig.~3(b) in the
main text. It can be seen that the topologically protected edge modes are stable up to disorder strength $W\simeq 0.5
w$, where $w$ is the static nearest neighbor hopping strength. Note that for class A model the gapless edge may not
cross at energy $\Omega/2$, in contrast to systems with particle-hole or chiral symmetries. 
\end{widetext}

\end{document}